\documentclass{jltp}
\usepackage{graphicx}

\runninghead{M. Niemetz and W. Schoepe}{On the Statistical Analysis of Single Vortex Nucleation Events}

\newcommand{\vHe}{$^4$He}

\newcommand{\ind}[1]{\mathrm{\sf#1}}
\newcommand{\vc}{\ensuremath{v_\ind{c}}}
\newcommand{\ivHe}{$^\mathit{\small 4}\!\!$He}
\newcommand{\vcn}{\ensuremath{v_\ind{c0}}}
\begin {document} 
\title{On the Statistical Analysis of Single Vortex Nucleation Events in Superfluid $^4$He}
\author{M. Niemetz and W. Schoepe}
\address{Institut f\"ur Experimentelle und Angewandte Physik,\\
Universit\"at Regensburg, D-93040 Regensburg, Germany}
\date{\today}
\maketitle
\begin {abstract}
We investigate the statistical properties of single phase slip events observed in vortex nucleation experiments in \ivHe\ by the groups in Berkeley and Paris. From the cumulative distribution function of the events we calculate the slip rate as a function of flow velocity. The critical velocity is defined as the mean velocity and its statistical width as the standard deviation. From the slip rate and from the observed linear temperature dependence of the critical velocity we obtain the energy barrier for vortex nucleation which is a quadratic  function of the flow velocity. A comparison with the statistical properties of the laminar to turbulent transition in the flow around an oscillating sphere shows strikingly different behaviour.

\end{abstract}

\section{Introduction}
Nucleation of individual quantized vortices in the flow of superfluid \vHe\ through micro-apertures has been observed since 1985.\cite{PB,p,b} In these beautiful and meanwhile famous experiments the $2\pi$ phase slip of the superfluid wavefunction is detected which occurs when a vortex crosses the stream lines of the aperture. The nucleation event was shown to be a stochastic process depending on the flow velocity and the temperature. Consequently, a statistical analysis of the events is performed. Quantities like the cumulative distribution function, slip rate, critical velocity and its statistical width are introduced and analyzed in order to gain insight into the physics of the nucleation process. One of the most interesting properties is the energy barrier which a vortex has to overcome during nucleation either by thermal activation or by quantum tunneling. The dependence of the barrier on the flow velocity is of particular importance for developing a physical model for the creation of vortices.

In our present work we are investigating the statistical properties of the nucleation events in a different way. The motivation comes from our own experiments on the transition from laminar flow to turbulence around an oscillating micro-sphere in superfluid \vHe\ at millikelvin temperatures.\cite{mn} The onset of turbulence in that case is a stochastic process which can be described appropriately by means of reliability theory. With this tool at hand we are re-analyzing the statistical properties of the single vortex nucleation events. Our method is simple, direct and transparent. It avoids mathematical shortcomings of the previous analysis by the authors. From the rigorous relation between cumulative distribution function and slip rate we find the latter as a function of flow velocity $v$. From this result we then infer the energy barrier $E(v)$ which is consistent with the observed linear temperature dependence of the critical flow 
velocity \vc. We are analyzing both ac and dc flow experiments and find similarities as well as differences. Finally, we compare our results of the phase slip experiments with the statistical analysis of our own experiment.

\section{Basics of Reliability Theory}
In this Section we present the basic tools of reliability theory\cite{gne} needed for the statistical analysis of the phase slip events. The cumulative distribution function (CDF) of the probability of observing a slip below a given velocity amplitude $v$ is $P(v)$, which is obtained from recording a time series of slip events. In order to employ the terminology of reliability theory we consider ``slip'' as ``failure'' and ``no-slip'' as ``reliability''. The reliability function
\[
R(v)\,=\,1\,-\,P(v)
\]
is determined by the failure rate $\Lambda (v)$:

\[
R(v)=\exp\left( - \int\limits_0^v \Lambda (v')\,dv' \right), 
\]
from which follows

\begin{equation}
\Lambda (v)=\, -\frac{d\,\ln R}{dv}\;.\label{one}
\end{equation}
We note that $\Lambda$ has the dimension of 1/velocity, i.e., s/m. It can be simply transformed to units of 1/s by the relation

\[\Lambda (t)=\Lambda (v)\,\frac{dv}{dt}\;,
\]
where $dv/dt$ is the change of the flow velocity amplitude with time. Because in the experiments the amplitude increases linearly, $dv/dt$ is a constant during the time series and depends only on drive and dissipation. The probability density function (PDF) $f(v)$  is given by
 
\[
f(v)=\frac{dP}{dv}=\,-\frac{dR}{dv}\;.
\]
From the experiment $f(v)$ is obtained if histograms are evaluated. The failure rate may be calculated directly from $f(v)$  by using

\[
\Lambda (v)=\frac{f(v)}{\int\limits_v^\infty f(v')\,dv'}\;.
\]
The critical velocity \vc\  of the flow may be defined as the mean velocity

\[
\vc\equiv \langle v \rangle = \int\limits_0^\infty v\cdot f(v)\,dv\;.
\]
Alternatively, the median velocity \ensuremath{v_\ind{m}} may be used which is given by $P(v_\ind{m}) = R(v_\ind{m})= 1/2$
or by
 
\[
\int\limits_{0}^{v_\ind{m}}f(v)\,dv=\frac{1}{2}\;.
\]
In the following it is irrelevant which choice is used. Finally, because the transition of $P$ from 0 to 1 occurs over a finite velocity interval, a ``width'' $\Delta$\vc\  may be defined. We use the standard deviation
  
\[
\Delta \vc\equiv \sqrt{\langle v^2\rangle -\langle v\rangle^2}
\]
i.e., the square-root of the variance. This choice is numerically much safer than the authors' definition of the inverse slope of $P(v)$ at $P=1/2$ which is given by $1/f(v_\ind{m})$ and therefore depends only on one point on the PDF which may easily be subject to errors.

\section{Data Analysis}
In order to apply the above theory to a recorded time series we can start either from the CDF or from the PDF. In the published literature these data can be found in two places, namely in the recent review by the Paris group\cite{p}  where data at 12 mK are shown, and in a publication by the Berkeley group\cite{b} where a PDF at 0.46 K and a CDF at 0.79 K are displayed. We have scanned these data and digitized them for analysis. 
\noindent\begin{figure}[tb!]
\noindent\centerline{\includegraphics[width=0.8\linewidth]{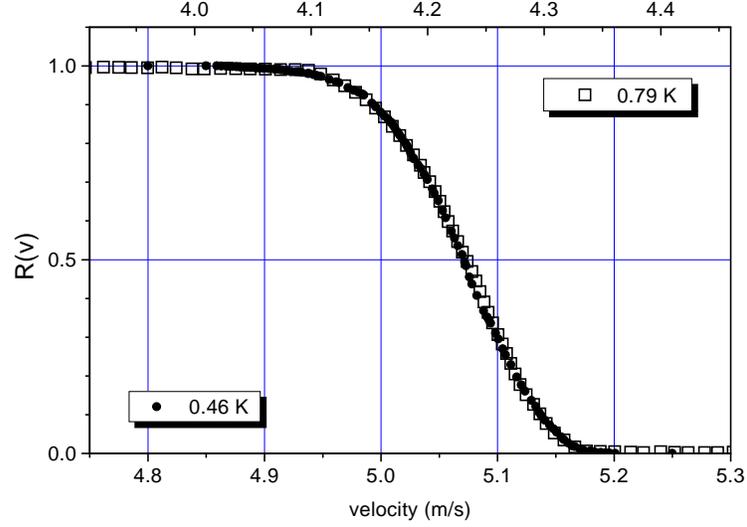}}
\caption{Reliability function $R(v)$ from the Berkeley PDF at 0.46\,K (bottom axis) and from the Berkeley CDF at 0.79\,K (top axis) (see Ref.\,\onlinecite{b}). Note the temperature independent shape. The overlap is obtained by shifting the velocity scales by 0.84\,m/s with respect to each other.}
\label{cdf}
\end{figure}\noindent%
\noindent\begin{figure}[tb!]
\noindent\centerline{\includegraphics[width=0.49\linewidth]{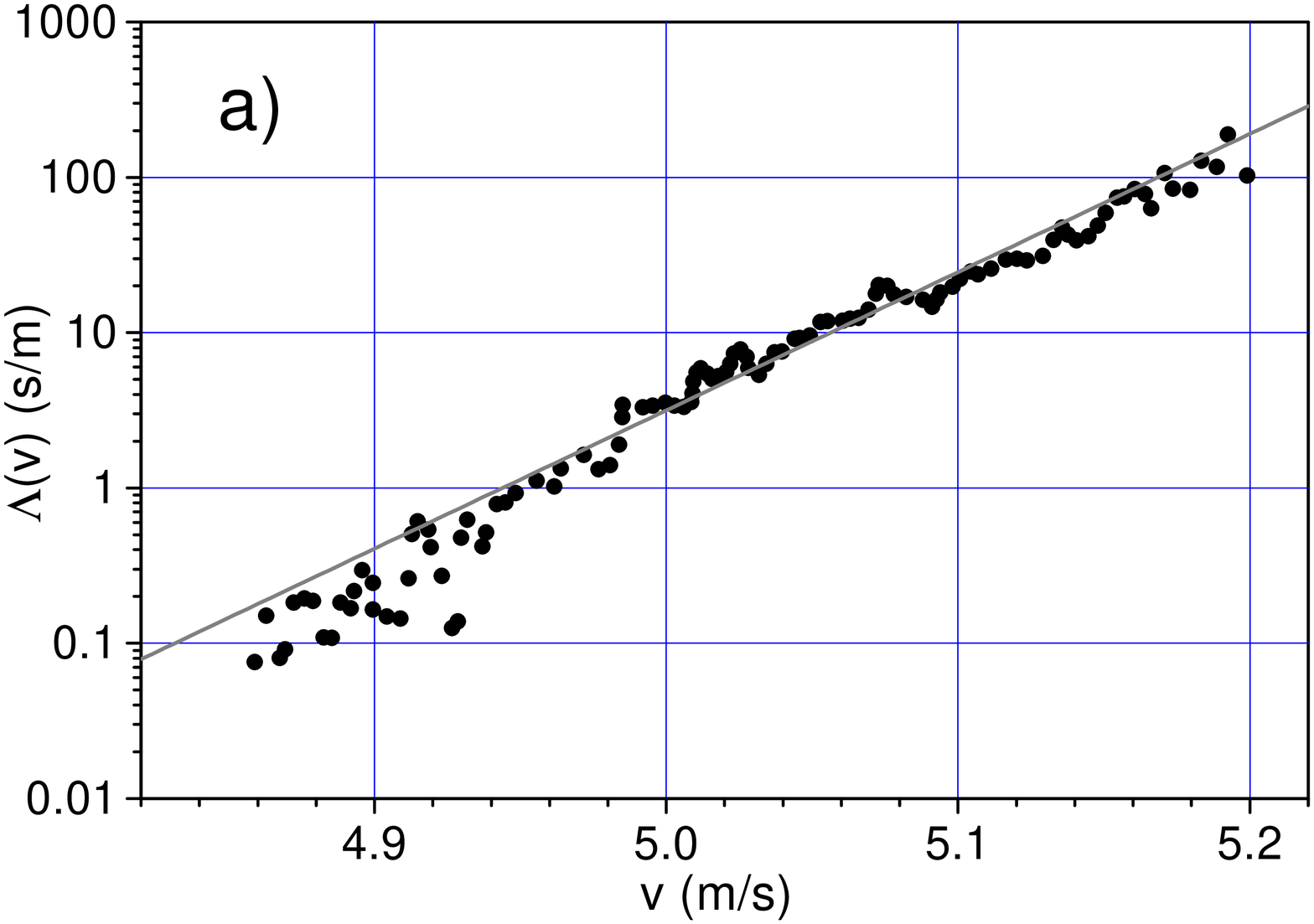}
\includegraphics[width=0.49\linewidth]{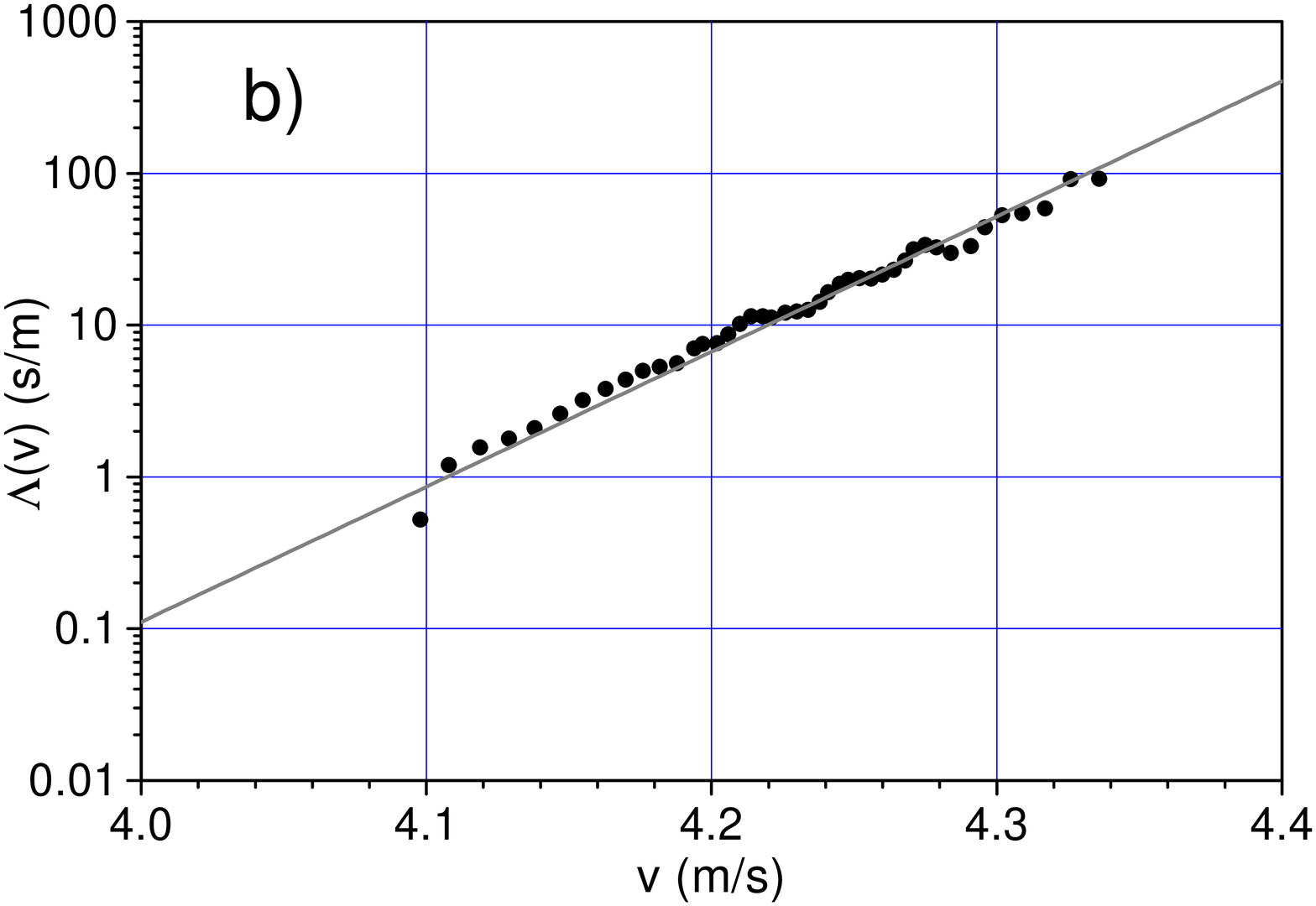}}
\caption{Slip rate $\Lambda (v)$ obtained from the data in Fig.\,\ref{cdf}: a) 0.46\,K, b) 0.79\,K. Note the exponential variation having the same slope for both temperatures.}
\label{lambda}
\end{figure}\noindent%

A first very interesting observation is the temperature independence of the shape of the Berkeley CDFs. Both curves overlap if the velocity scale is shifted, see Fig.\,\ref{cdf}. This shift is due to a temperature dependence of \vc\ (or $v_\ind{m}$) which is linear.\cite{b} At 0.46 K and at 0.79 K we find \vc\ = 5.07 m/s and 4.23 m/s, respectively. The median velocity \ensuremath{v_\ind{m}} is larger by only 0.01 m/s at either temperature. The constant shape implies a constant width $\Delta$\vc\ = 0.0611 m/s. The slip rate $\Lambda (v)$ is obtained by numerical differentiation, see Eq.\,(\ref{one}). The results for both temperatures are shown in Fig.\,\ref{lambda}. An exponential increase of $\Lambda $ is obvious with a temperature independent slope~$b$\,:
\[
\Lambda (v,T) = a(T)\cdot\exp(b\cdot v) \;,
\]
where $b$ = 20.5 s/m ($1/b=0.0488$\,m/s) and $a(T)$ is a temperature dependent coefficient. Inspection of $a(T)$ gives the following surprisingly simple relation:
\[a(T) = \Gamma _\ind{v}\cdot \exp(-b\cdot \vc\ (T)),
\]
where $\Gamma _\ind{v}$ = 14.3 s/m. Therefore, we have finally
\begin{equation}
\Lambda (v,T) = \Gamma _\ind{v}\cdot \exp\{b\cdot (v - \vc(T))\},\label{two}
\end{equation}
and at all $v =  \vc(T)$ we obtain $\Lambda (\vc) = \Gamma _\ind{v}$. The critical velocity is the  ``working point'': because of the rapid variation of $\Lambda (v)$ data can be taken only in the vicinity of \vc. The time resolution of the experiment is limited by the half period of the oscillating flow because the amplitude of each half period is recorded. Therefore, the highest measurable slip rate in the Berkeley ac experiments is given by
$\Lambda (t) = 2\,f$, where $f = 40.7$ Hz is the oscillation frequency, or $\Lambda (v) = \Lambda (t)/ (dv/dt)$ = 581 s/m where we used the value $dv/dt$ = 0.14 m/s$^2$   obtained from Fig.\,\ref{shifted} of Ref.\,\onlinecite{b}. In fact, the original data appear to extend up to about this value. Because of the digitizing procedure our data are limited to about 200 s/m.

Performing the same analysis with the Paris CDF\cite{p} we obtain the same exponential behaviour of $\Lambda (v)$. Because in that work the velocity is given in dimensionless units, we cannot compare the absolute values directly with the above results from the Berkeley data. We find: \vc = 59.64, $\Delta $\vc\ = 0.195, $b$ = 6.23, $\Gamma _\ind{v}$ = 3.4. The highest slip rate (measured at $v$ = 60) is $\Lambda (60)$ = 15 and from $dv/dt \approx 2.1$  s$^{-1}$ (from Fig.\,\ref{shifted} of Ref.\,\onlinecite{p}) we get the maximum rate $\Lambda (t)$ = 32 s$^{-1}$ which corresponds to one slip per half period (which is 31.8 ms).\cite{p1}

From the expression of the slip rate (see Eq.(\ref{two})) we easily obtain an analytical form of $R(v)$ and $f(v)$. Neglecting the small difference between \vc\  and \ensuremath {v_\ind{m}}  we find from $R(v_\ind{m}) =1/2$ the following results: $f(v_\ind{m}) = \Gamma_\ind{v} /2$ and $\Gamma_\ind{v} = b\cdot \ln  2$. Both the Berkeley and the Paris data are consistent with this. It is quite plausible that only two parameters are left to determine the properties of the CDF or the PDF: the critical velocity \vc($T$) , which determines the position on the velocity axis, and the width $\Delta $\vc , which can be expressed either by $\Gamma_\ind{v}$ or by $b$. 

We now draw conclusions from the temperature independence of the slope $b$. Because the ac data can be taken only in the vicinity of \vc\ it is reasonable to compare Eq.\,(\ref{two}) with a Taylor expansion of the energy barrier $E(v)$ for vortex nucleation which has to be overcome by thermal activation:\cite{b}

\[
E(v) - E(\vc) = \left(\frac{dE}{dv}\right) _{\vc}\cdot(v-\vc ).
\]

Comparing $-(E(v)-E(\vc ))/T $ with the exponent in Eq.\,(\ref{two}) we find
\[
b = -\frac{1}{T}\left(\frac{dE}{dv}\right) _{\vc (T)}
\]
which according to the experiment must be independent of $T$. This implies a differential equation
\[
\left(\frac{dE}{dv}\right) _{\vc (T)}= -b\,T
\]
where $T$ is the inverse function of $\vc(T)= \vcn(1-T/T_0)$ which is obtained from the experiment. Integration then gives
\[
E(v) = E_0 - T_0\,b\,v + \frac{T_0\,b}{2\vcn }v^2 
\]
for all $v=\vc (T)$, i.e., for $0\leq v\leq \vcn $. Choosing $E(\vcn )=0$ fixes the integration constant to the value $E_0=b\,\vcn\ T_0/2$ and gives
\begin{equation}
E(v)=E_0\left(1-\frac{v}{\vcn }\right)^2.\label{three}
\end{equation}
Inserting $b$=20.5 s/m, \vcn =6.24 m/s and $T_0$= 2.45 K we get $E_0$= 157 K. We thus have the important result that the energy barrier is reduced by the flow velocity in a parabolic form. We emphasize that this is the consequence of the linear variation of \vc ($T$) and the temperature independence of the slope $b$ of the slip rate. If $b$ becomes temperature dependent the energy barrier as a function of velocity will contain higher order terms. 

\begin{figure}[tb!]
\noindent\centerline{\includegraphics[width=0.8\linewidth]{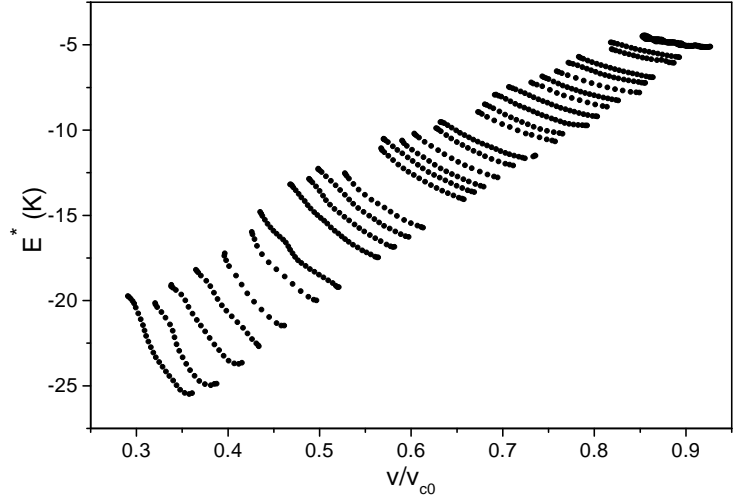}}
\caption{Results for the energy barrier $E^*$ versus dc flow velocity at various temperatures ranging from 1.83\,K (lower left) to 0.35\,K (upper right) obtained by the Berkeley group (digitized data from Fig.\,10 of Ref.\,\onlinecite{b}).}
\label{unverschoben}
\end{figure}

\begin{figure}[tb!]
\noindent\centerline{\includegraphics[width=0.8\linewidth]{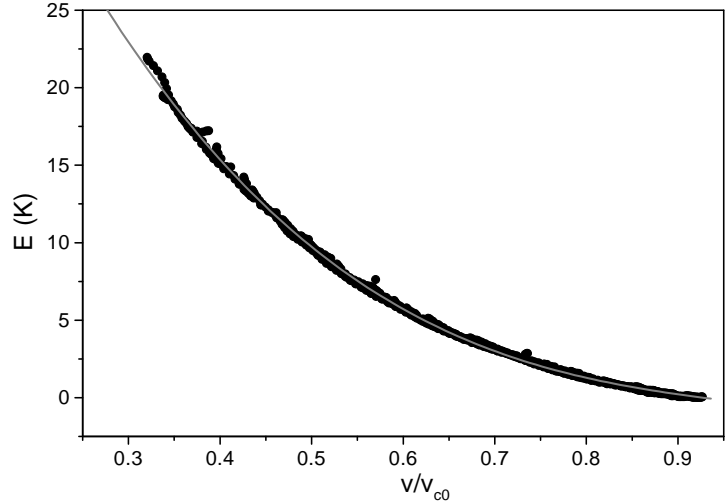}}
\caption{Connecting all isotherms of Fig.\,\ref{unverschoben} by vertical shifts produces an universal velocity dependence of the energy barrier. The fitted curve is a third order polynomial, see text. Energy zero chosen at $v/\vcn =1$.}
\label{shifted}
\end{figure}

We are now in the position to extend the analysis of the dc flow experiments of the Berkeley group (Fig.\,10 of Ref.\,\onlinecite{b}). In these experiments the energy barrier $E^*$ is obtained at various constant temperatures for a certain range of velocities, where $E^*$ is defined as
\[
E^*(v,T) = E(v) - T\cdot \ln \Gamma. 
\] 
$E(v)$ is the kinetic energy barrier and $\Gamma $ is an ``attempt frequency'' (in Hz), see Fig.\,\ref{unverschoben}. Using the same energy scale as with the slip rate, namely  $E(v)-E(\vc(T) )$, we calculate the energy difference $\Delta E^*$ between two isotherms $T_1$ and $T_2$ at the same velocity $v$ to be
\[
\Delta E^* = -E(\vc (T_1))+E(\vc (T_2))-(T_1-T_2)\ln\Gamma 
\]
where $\Gamma$ is assumed to be constant in temperature. Inserting $\vc (T)$ into Eq.\,(\ref{three}) gives $E(\vc (T))= E_0T^2/T_0^2$ and hence
\[
\Delta E^*=-E_0(T_1^2-T_2^2)/T_0^2-(T_1-T_2)\ln \Gamma 
\]
Introducing $\Delta T=T_2-T_1$ as a new variable leads to
\begin{equation}
\Delta E^* = -\frac{E_0}{T_0^2}(\Delta T)^2 + \left(\frac{2T_2E_0}{T_0^2}+\ln \Gamma \right)\Delta T.\label{four}
\end{equation}
\noindent\noindent%
\noindent\noindent%

\begin{figure}[tb!]
\noindent\centerline{\includegraphics[width=0.8\linewidth]{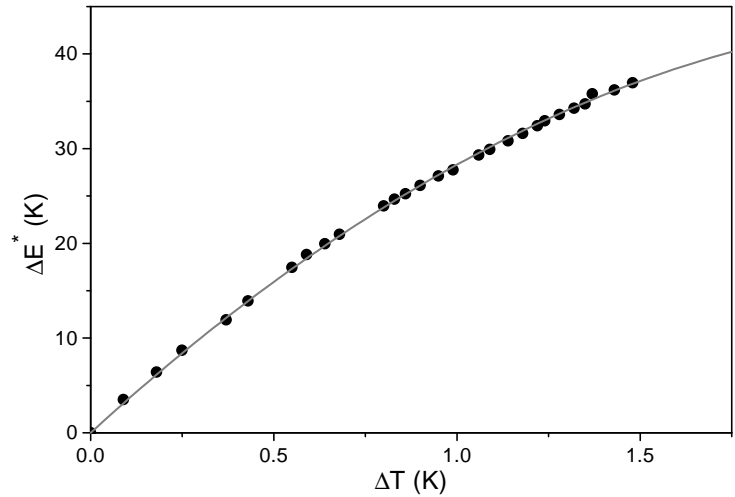}}
\caption{Energy shifts $\Delta E^*$ obtained from constructing the connected curve of Fig.\,\ref{shifted} versus the temperature difference from 1.83\,K. Equation (4) is fitted to the data.}
\label{civonti}
\end{figure}

We construct now the universal $E(v)$ curve for all dc data by connecting succesively all measured isotherms in Fig.\,\ref{unverschoben} by a vertical shift to the data at the reference isotherm $T_2=1.83$\,K, see Fig.\,\ref{shifted}. The resulting curve follows approximately the quadratic variation of Eq.\,(\ref{three}), but a small third order term (15\% at the highest $E^*$ values) is needed for a perfect fit. In Fig.\,\ref{civonti} the vertical shifts $\Delta E^*$ are shown and Eq.\,(\ref{four}) is fitted. This yields $E_0=57$\,K (for $T_0=2.83$\,K extrapolated from Fig.\,9 of Ref.\,\onlinecite{b}) and $\Gamma =1.2\cdot10^4$\,Hz. Although these values differ from those of the ac experiments discussed above the general properties of both experimental results are similar.
\noindent\noindent%

\section{Conclusion}
Starting from the correct relation between cumulative distribution function and slip rate we have re-analyzed the published data on single phase slip events. We find a slip rate which varies exponentially with flow velocity. We deduce the energy barrier for vortex nucleation, decreasing quadratically with flow velocity. Finally, we extend the analysis of the dc flow experiments and obtain a consistent picture for the results of both methods. We do not offer any physical model for the results of our analysis. Further work on vortex nucleation is needed to explain the linear temperature dependence of the critical velocity, the exponential increase of the slip rate or the quadratic dependence of the energy barrier on the flow velocity.

Returning to our own experiment on the transition to turbulence in the oscillatory flow of superfluid \vHe\  at millikelvin temperatures\cite{mn} we note that in our case the failure rate (in units of s/m) of the laminar phase is proportional to the velocity increase above the critical velocity for turbulent flow. The lifetime (in units of s), however, may become metastable being limited ultimately only by creation of vorticity due to natural background radioactivity or cosmic rays. The lifetime of the turbulent phase grows rapidly with increasing drive power, becoming infinite at some critical value. The stability of both the laminar and the turbulent phases can probably not be understood in terms of single vortex nucleation effects but rather by the dynamics of fully developed superfluid turbulence.               

\section*{Acknowledgment}
We are very grateful to A. Kalbeck and V. Nov\'ak for advice on digitizing figures and to \'E. Varoquaux for sending a preprint of Ref.\,\onlinecite{p}. This work is supported by the Deutsche Forschungsgemeinschaft.

\end {document}